# Sequential Electronic and Structural Transitions in VO$_2$ Observed Using X-ray Absorption Spectromicroscopy

*Suhas Kumar[1,2*], John Paul Strachan[1], Matthew D. Pickett[1], Alexander Bratkovsky[1], Yoshio Nishi[2], and R. Stanley Williams[1]*

[1]*Hewlett-Packard Laboratories, 1501 Page Mill Rd, Palo Alto, CA 94304, USA*
[2]*Stanford University, Stanford, CA 94305, USA*
E-mail: suhaskumar@stanford.edu

**The popular dual electronic and structural transitions in VO$_2$** are explored using x-ray absorption spectromicroscopy with high spatial and spectral resolutions. It is found that during both heating and cooling, the electronic transition always precedes the structural Peierls transition. Between the two transitions, there are intermediate states that are spectrally isolated here.

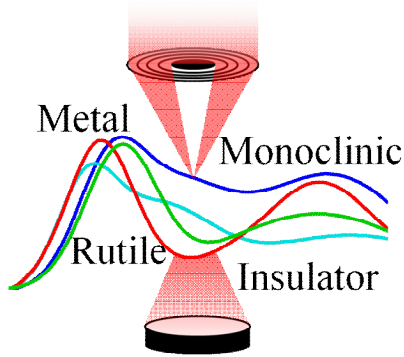

Vanadium dioxide (VO$_2$) undergoes a hysteretic metal-insulator transition (MIT) and a structural phase transition (SPT) from monoclinic to rutile phases, upon heating or cooling through a temperature range near 340 K.[1] MIT and SPT are receiving renewed interest because of potential applications in non-volatile memory, optics, micromechanics and neuromorphic computing.[2-5] Recent experimental and theoretical studies[1,6-8] have still not resolved significant questions about the nature of the transitions that have lingered for nearly half a century[1,9-10] regarding the relationship between the transitions and their sequence of occurrence,[11-15] the existence of intermediate states,[1,11,15] and the nature of the insulating state.[8,16] We performed x-ray absorption spectromicroscopy to monitor the π*[17-18] and d$_∥$* conduction bands,[1,19] which have been previously associated with the MIT and SPT, respectively,[6,11,20-21] while simultaneously measuring in-situ the macroscopic resistance of a temperature-ramped thin film of VO$_2$. During the transitions between the limiting phases, namely, monoclinic-insulator and rutile-metal, we observed the occurrence of intermediate spectra in sub-micrometer domains which could not be assigned to a simple combination of either, and instead we assigned them to a rutile insulator on cooling and a monoclinic metal on heating. The designation of insulator or metal was determined by the spectral position of the π* band and corroborated by the simultaneous resistance measurement. The sequential transitions have been inferred before through several techniques, but distinct intermediate states were not isolated.[8,11,16,19,22] This observation required the use of a technique that could simultaneously distinguish both the MIT and SPT with high spatial and spectral resolutions of <100 nm and <100 meV, respectively.

We grew 40 nm of $VO_2$ by oxidizing a thin film of vanadium[23] on a 20 nm silicon nitride membrane to enable x-ray transmission experiments. Platinum electrodes were subsequently deposited with a titanium adhesion layer to simultaneously monitor film resistance across 4 µm of the oxide. The sample was mounted on a holder that was attached to a heater and the film temperature was monitored with a thermocouple placed near the area under measurement (**Figure 1a**). The film exhibited a hysteretic resistance change during a heating-cooling cycle across the transition regime (Figure 1b). The scanning transmission x-ray microscope (STXM) used highly focused x-rays with less than 30 nm spot diameter and tunable wavelength with 70 meV energy resolution for soft x-ray absorption spectroscopy (XAS).[24] We used the oxygen-K edge absorption to study the unoccupied bands, specifically the $\pi^*$ and $d_\parallel^*$ peaks (both containing the 3d states of vanadium). Movement of the π* band to lower energies indicated the development of states at the Fermi energy and metallic behavior, while the formation of a $d_\parallel^*$ peak revealed V-V dimerization in a monoclinic structure, which is a Peierls transition.[17-19] By analyzing the phase composition of the film with temperature, we confirmed that the spectral position of the π* band almost completely accounts for the resistance behavior. In this way, we were able to separately monitor MIT and SPT using spatially resolved XAS.

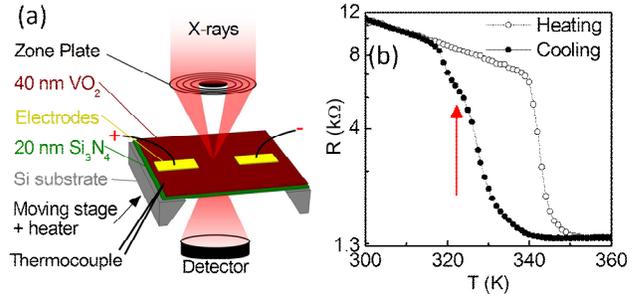

**Figure 1.** (a) Schematic of the setup. (b) Resistance of the film with ambient temperature showing hysteresis during a heating-cooling cycle. The red arrow indicates a shoulder in the cooling curve, referred to in the text.

XAS spectra were collected along a 10 µm line at steps of 67 nm during heating and cooling cycles with film temperature ranging from 300 to 357.1 K. Figure 2a shows a color map of some representative spectra. These spectra show the three identifiable bands corresponding to $VO_2$: $\pi^*$, $d_\parallel^*$ and $\sigma^*$.[1] The limiting spectra (at 300 K and 357.1 K in Figure 2a, also shown in red and blue in Figure 2b) have been assigned previously to the monoclinic insulator (M-insulator) at low temperatures and rutile metal (R-metal) at high temperatures.[17-18,25] From the raw line-scan spectra in Figure 2a, we see that an energy shift of the π* band always precedes the appearance or disappearance of the $d_\parallel^*$ peak in several spatially localized domains. By further examining the spectra in spatially localized portions of the line scan, we identified two consistent intermediate states. In heating from M-insulator, we found a state for which the π* band had shifted down to indicate metallic conductivity but the $d_\parallel^*$ band was still present (Figure 2b), which we assign to a monoclinic metal (M-metal) phase. Such a state has been predicted previously, but not spectrally isolated during the transition.[6,11,14,19-20,25-30] On cooling from rutile metal (R-metal), we found a different intermediate state in which the π* band had shifted upward to indicate an insulator but no $d_\parallel^*$ band had appeared, which we assign as a rutile insulator (R-insulator). The four spectra shown in Figure 2b are averaged over multiple domains of the same state, and a spatial mapping of these states is shown in Figure 2c for both heating and cooling. The maps display spatially localized evolution and statistical significance of the intermediate states. We emphasize that these are distinct spatially localized intermediate states, for which the spectral signatures cannot be reproduced by using a combination of spectra from the limiting states.

To summarize our observations, we plot the fraction of each of the phases as a function of temperature during heating and cooling in Figure 3a. The proportion of the intermediate phases (green and cyan) peaks at the centroid of the resistance transition and then disappears at the limiting phases. Separately, at each temperature, we decomposed the spectra to track the $\pi^*$ energy shift and presence of the $d_\parallel^*$ band. This provided a phase fraction for the insulating and monoclinic phases separately. We then compared these fractions to the corresponding resistance value at the same temperature. Shown in Figure 3b is the comparison of the resistance vs. phase fraction behavior to that predicted by the Bruggeman effective medium approximation (EMA) for percolating systems, used to compute the macroscopic resistance given the volume fraction of the component phases with differing conductivities.[31-33]

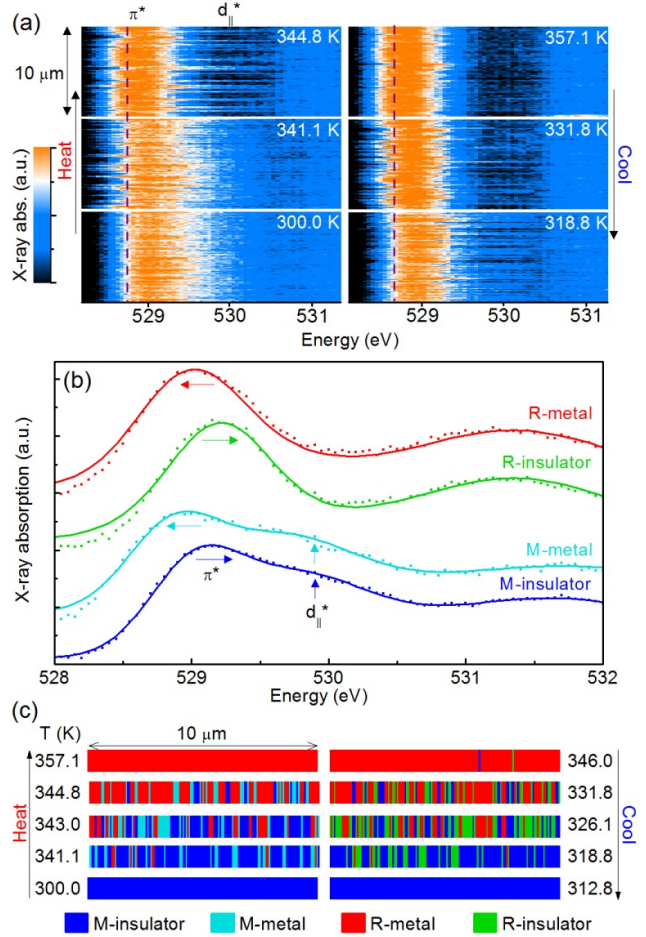

**Figure 2.** (a) Line-scan spectra are shown for three different temperatures on both the heating and cooling branches. These show that the transition of the $\pi^*$ band energy in several domains always precedes the appearance or disappearance of the $d_\parallel^*$ band (indicated by black for low intensity or the absence of the band). Dashed purple lines are an aid to distinguish the leading edge of the $\pi^*$ band (orange) in the metallic and insulating states. (b) Distinct, isolated O K-edge spectra corresponding to the four states during the transition. Those of the intermediate states were isolated from the line-scans by averaging over nine domains with similar spectra. (c) 1-D maps of the film during a heating-cooling cycle across the transition. O K-edge spectra were obtained at 150 points along a 10 μm line. Spectra from (b) were used to identify the states of the different individual spatial regions.

We see that the film resistance quantitatively tracks that predicted by the fraction of the insulating phases ($\pi^*$), while it diverges from that predicted by the fraction of the monoclinic phases ($d_\parallel$*). This reinforces the fact that the energy of the lowest conduction band determines the resistance during the MIT in VO$_2$. For cooling, we note that the fraction of the intermediate phase reaches its maximum at about 326.1 K, where the corresponding resistance curve of Figure 1b develops a shoulder (indicated by a red arrow), which is a corroboration of the stabilization of the intermediate state (also seen in Figure 2c) and its effect on the resistance.

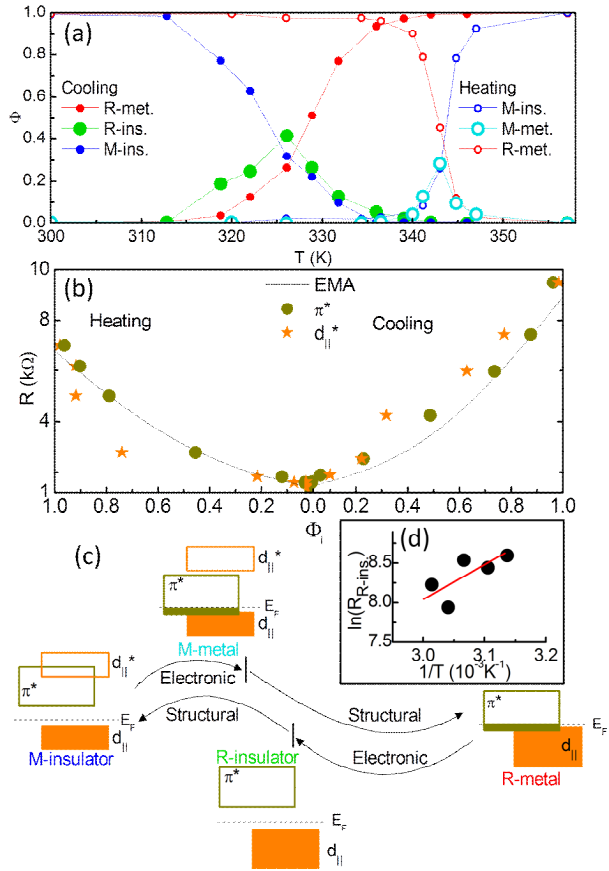

**Figure 3.** (a) Fraction of the component states, ϕ, during a heating-cooling cycle across the transition. The intermediate states show up prominently during the transitions, but are not present outside the transitions. (b) Comparison of the prediction of resistance behavior with phase fraction, ϕ, obtained from effective medium analysis and the phase fractions of insulating and monoclinic phases from panel (a) and Figure 1b. (c) A schematic of the transition drawn from the analysis in Figure 2-3b. Sequential electronic and structural transitions during heating and cooling allows for distinct intermediate states to exist during both. Signatures of these intermediate states were observed in Figure 2, while the sequential transitions were inferred from Figure 2c and 3a. As noted in the text, we crudely estimated the bandgap for the M-insulator to be 0.3-0.4 eV and that for the R-insulator to be ~0.4 eV. (d) Plot of the natural logarithm of resistance of the rutile insulator phase with inverse temperature. Red line is a least-square linear fit to data.

We qualitatively describe the transitions using the schematic diagram in Figure 3c. The band structures for the M-insulator and R-metal states are similar to the model of Goodenough,[17] while those for the two intermediate states involve shifting the π* conduction band up for the M-insulator or down for the R-metal phases. From the resistance data, we calculated the activation energy of the conductivity of the monoclinic insulator to be 0.350±0.002 eV. The shift in the π* band was observed to be 0.15-0.2 eV to higher energy compared to the metallic states (limited by the energy resolution of the x-ray monochromator). If there is a corresponding downshift of the valence band to lower energies,[34] we can obtain a crude estimate of 0.3-0.4 eV for the bandgap of the monoclinic insulator, in agreement with the electronic activation energy. The M-metal state has the π* band shifted to lower energies to touch the Fermi level, whereas the R-insulator state has a gap between the Fermi level and the π* band but no antibonding V-V dimer states. In both the rutile states, we assumed a completely filled $d_\parallel$ valence band, mainly because we can completely account for the unfilled conduction bands with only the π* band. The MIT from R-metal to R-insulator on cooling is mainly electronic, and is followed by a structural transition from R-insulator to M-insulator. On heating, the first transition is again electronic, M-insulator to M-metal, and the second transition is structural, M-metal to R-metal. Thus we see that in both cases, a predominantly Mott-like transition precedes a Peierls transition. The cooling and heating ramps transition through different intermediate states. A purely electronic transition will likely have a smaller activation energy than a structural transition,[20] so for this system the electronic transition always precedes the Peierls transition. However, the energies of all four phases are extremely close to each other, so that very subtle effects can determine the ground state and the transitions among the phases.

The R-insulator phase is interesting since it may be responsible for the shoulder on the resistance vs. temperature cooling curve in Figure 1b, which coincides with the maximum in volume fraction of this phase shown in Figure 3a. In an attempt to determine the electrical properties of the R-insulator, we used a multi-phase percolation model[33] to determine the resistance and plot the temperature dependence of this state (Figure 3d). By assuming activated transport, we estimated the activation energy to be 0.4±0.2 eV, which has a large uncertainty because of the narrow temperature window for which data were available. Given that the upward shift of the R-insulator $\pi^*$ state is slightly larger than that of the M-insulator and again assuming a corresponding downshift of the valence band to lower energies,[34] we obtained a crude estimate of 0.4 eV for the bandgap of the rutile insulator, in agreement with the resistance activation energy for this state.

Previous work[35] has determined that a complete picture of the phase transitions in single-crystalline and disordered $VO_2$ structures requires both ferroelastic and metal-insulator strain effects to be taken into account. There have also been demonstrations of reversal in the sequence of transitions or suppression of a transition caused by external stimuli like strain, doping, etc.[7,21] Recent work also showed a solid state triple point in $VO_2$ in a stress-temperature phase diagram, although the intermediate states described here were not observed.[36] These states have not been directly observed previously because it is necessary to have an experimental probe with both excellent spatial resolution (large area probes average over multiple domains of different phases) and the capability to simultaneously detect the structural and electrical properties of the material in a single localized domain. In addition, collecting resistance data during the spectromicroscopic analysis was key to understanding the transitions. We were able to isolate the intermediate phases by working with a sample that had a broad transition region of ~40 K and using a stable temperature stage with good resolution. The sequence of the transitions exhibited here and the intermediate states have been proposed previously from both theory and experiments,[8,11,16,19,22,34] but because of the sensitivity of the $VO_2$ system to strain and composition, the transitions could have a different sequence and/or intermediate states depending on sample growth parameters and experimental measurement conditions.[1,7,11,15,21,37]

In conclusion, observations of the MIT and SPT in thin film $VO_2$ were performed by x-ray absorption spectromicroscopy during a heating-cooling cycle. The two transitions were observed to occur sequentially, leading to the emergence of stable intermediate states possessing distinct spectroscopic signatures. We showed the importance of using techniques with high spectral, spatial and temperature resolution to accurately study inhomogeneous media, which may have complicated multi-dimensional phase diagrams.

**Note added post-proof:** Two other groups reported very similar results around the same time this article was published: *Phys. Rev. Lett. 113, 216401 (2014); Phys. Rev. Lett. 113, 216402 (2014)*

**Acknowledgements**
All x-ray measurements were performed at the Advanced Light Source in Lawrence Berkeley National Laboratory, Berkeley, CA, USA. The Advanced Light Source is supported by the Director, Office of Science, Office of Basic Energy Sciences, of the U.S. Department of Energy under Contract No. DE-AC02–05CH11231.